\documentclass[prl,twocolumn,showpacs,amsmath,superscriptaddress]{revtex4}

\usepackage[dvips]{graphicx}
\usepackage{graphics}
\usepackage{dcolumn}
\usepackage{bm}

\begin{document}

\title{Relation between Hall resistance and the diamagnetic moment of
Fermi electrons}

\author{P. St\v{r}eda}
\affiliation{Institute of Physics, Academy of Sciences of the
Czech Republic, Cukrovarnick\'{a} 10, CZ - 162 53 Praha}

\date{\today}

\begin{abstract}
General thermodynamical arguments are used to relate the Hall
current to the part of the magnetic moment originated in 
"macroscopic current loops". The Hall resistance is found
to depend only on the electron properties in the vicinity
of the Fermi energy, which is the essential advantage of
the presented treatment. The obtained relation is analyzed
by using Landauer-B\"{u}ttiker-like view to the electron
transport. As one of the possible application the Hall
resistance of the periodically modulated two-dimensional
electron system in strong magnetic fields is briefly discussed.
\end{abstract}


\maketitle

Relation between Hall effect and the diamagnetic moment of
carriers has been studied for decades and often it has been
the subject of the controversial discussion. The basic
thermodynamical arguments are trivial. The diamagnetic moment
per unit volume $\vec{M}$ is defined by the first derivative
of the grand canonical potential $\Omega$ with respect of the
magnetic field $B$
\begin{eqnarray} \label{Mdef}
\vec{M} \, = \, - \frac{\vec{B}}{B}
\left( \frac{\partial \Omega}{\partial B} \right)_{\mu} \\
d\Omega(\mu,B) \, = \, - \, \vec{M} d\vec{B} \, - \,
N d\mu \; ,
\end{eqnarray}
where $N$ and $\mu$ are carrier concentration and chemical
potential, respectively. Magnetic moment is supposed to be
parallel with the magnetic field direction and the zero
temperature is considered for simplicity. Identifying
$c \, {\rm curl} \, \vec{M}$ with a current the linear response
to the applied electric field
$\vec{\cal{E}} \equiv \vec{\nabla} \mu /e$ immediately gives
\begin{equation} \label{Jdef}
\vec{J} \, = \, - ec
\left( \frac{\partial \vec{M}}{\partial \mu} \right)_{\vec{B}}
\times \vec{\cal{E}} \; ,
\end{equation}
where $-e$ ($e>0$) denotes the electron charge. Since the resulting
current $\vec{J}$ is perpendicular to the  applied electric field
the tendency to identify it with the Hall current \cite{Widom} often
appears. However, it is not so simple as has been already discussed
e.g. by Hajdu \cite{Hajdu}. Nevertheless, it is generally accepted
that at least surface diamagnetic currents have to be taken into
account to obtain correct values of the magnetization and that
they also substantially influence the non-diagonal components
of transport coefficients
\cite{Obraztsov,Zyrjanov,Peletminskii,Smrcka,Streda84}.
For the sake of the simplicity the following treatment is
limited to two-dimensional electron systems in perpendicularly
applied magnetic fields. 

It has been proved by several different methods that the direct
relation between Hall current and magnetization, Eq.~(\ref{Jdef}),
can be applied whenever the chemical potential $\mu$ is located
within the energy gap of the electron energy spectrum
\cite{Streda82,Streda83,Rammal,Schenker}. For periodic systems
the Maxwell relation
\begin{equation} \label{Maxwell}
\left( \frac{\partial M}{\partial \mu}\right)_{\vec{B}}
\, \equiv \,
\left( \frac{\partial N}{\partial B}\right)_{\mu}
\end{equation}
together with the condition that the magnetic flux per unit cell
$\Phi$ is a rational multiple of the flux quantum
$\Phi_0 \equiv hc/e$, gives quantized values of the
Hall conductance
\begin{equation} \label{QHE}
\sigma_{Q} \, = \,  - ec
\left( \frac{\partial N}{\partial B}\right)_{\mu}
\, = \, -\frac{e^2}{h} \, i \quad ; \quad
{\rm for} \; \mu \; {\rm in} \; {\rm gap} \; .
\end{equation}
Integer $i$ has to satisfy Diophantine equation
\cite{Wannier,Thouless,Streda94}
\begin{equation} \label{Diophantine}
\nu \, \equiv \, \frac{N}{\frac{eB}{hc}} \, = \,
i + s \frac{q}{p} \quad ; \quad 
\frac{p}{q} \, \equiv \, \frac{\Phi}{\Phi_0} \; ,
\end{equation}
where $\nu=N/(eB/hc)$ is the filling factor and integer $s$
is additional gap quantum number \cite{Streda94}.  For zero
modulation $s=0$, there are just $i$ fully occupied Landau
levels bellow the Fermi energy and the expression Eq.~(\ref{QHE})
represents integer quantum Hall effect \cite{Klitzing}.
Only recently the predicted non-trivial sequence of quantum
Hall values in periodically modulated systems has been
observed \cite{Geisler}.

The diamagnetic moment can alternatively
be evaluated by using expectation values of the corresponding
operator, i.e.
\begin{equation} \label{M-expect}
\vec{M}
\, \equiv \, - \frac{e}{2c} \, 
{\rm Tr} \left[ \Theta(\mu-H) \, \vec{r} \times \vec{v} \right]
\; ,
\end{equation}
where $\vec{r}$ and $\vec{v}$ are electron coordinate and
velocity operators, respectively, $\Theta(\mu-E)$ is the
Heaviside step function.
Using the Landau gauge for the vector
potential, $\vec{A} \equiv (0,xB,0)$, the single-electron
Hamiltonian representing a two-dimensional electron system
in the perpendicular magnetic field, say in $\hat{z}$
direction, obeys the following form
\begin{equation} \label{hamiltonian}
H \, = \, \frac{p_x^2}{2 m^{\ast}} +
\frac{ \left( p_y+\frac{eB}{c} x \right)^2}{2 m^{\ast}}   
+ V_{\rm b}(x,y) + V_{\rm conf}(x) \, ,
\end{equation}
where the confining potential $V_{\rm conf}(x)$ defines
width of the strip and $V_{\rm b}(x,y)$ is a background
potential. Periodic boundary conditions applied along $\hat{y}$
direction allow to describe energy spectrum in the form of branches
$\varepsilon_{\beta}(k)$. Each of the states given by a branch
index $\beta$ and the wave number $k$ has its own center of
the mass, $X_{\beta}(k)$, defined as the expectation value of
the $x$ coordinate. Because of the infinite strip length
the expectation values of the operator $-y v_x$ entering the
expression Eq.~(\ref{M-expect}) are not well defined.
Nevertheless, as the direct consequence of the current
conservation law it can be proved, that it gives the
same contribution to the magnetic moment as the operator
$x v_y$.

In order to find relation between Hall current and the magnetic
moment in the general case, let us decompose
the magnetic moment into two parts
\begin{equation}
\vec{M} \, \equiv \, \vec{M}_{i} \, + \, \vec{M}_{a}
\; .
\end{equation}
The part denoted as $\vec{M}_{i}$ originates in
microscopic current loops determined by the relative particle
motion with respect of the center of the mass. The remaining part
$\vec{M}_a$ appears due to the presence of the macroscopic currents
and it is defined as follows
\begin{eqnarray} \label{Ma}
M_{a} \, = \, - \frac{e}{c} \, \sum_{\beta,k}
\Theta \left( \mu - \varepsilon_{\beta}(k) \right) \,
X_{\beta}(k) \, v_{\beta}(k) \; , \\
v_{\beta}(k) \, = \, \frac{1}{\hbar} \,
\frac{d \varepsilon_{\beta}(k)}{dk} \; ,  
\end{eqnarray}
where $v_{\beta}(k)$ denotes velocity expectation values.

\begin{figure}[h]
\includegraphics[angle=0,width=2.5 in]{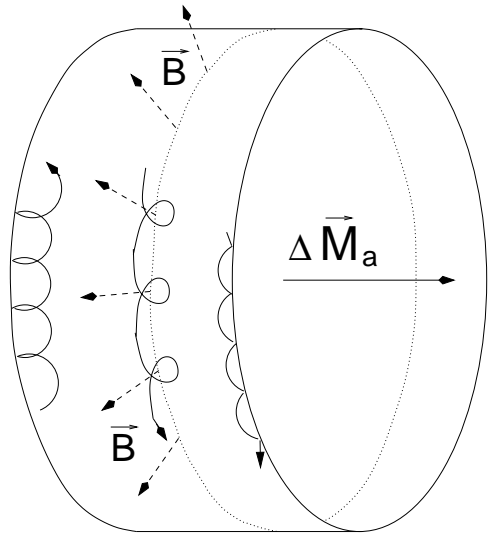}
\caption{Schematic view to the hollow cylinder formed from the two
dimensional strip. Magnetic field $\vec{B}$ is perpendicular to the
strip. The magnetic moment $\Delta \vec{M}_a$, parallel to the 
cylinder axis, represents the effect of the  macroscopic current   
loops while the magnetic moment $\vec{M}_i$ due to microscopic
currents is parallel to the local magnetic field.
On the edges skipping-like orbits representing edge states   
are sketched.}
\label{fig1}
\end{figure}

To clear up the role of the both parts, $\vec{M}_{i}$ and
$\vec{M}_{a}$, let us consider a hollow cylinder formed from
the strip as has been already suggested by Laughlin to
explain integer quantum Hall effect \cite{Laughlin}.
As illustrated in Fig.~1, the part $\vec{M}_{i}$
remains perpendicular to the strip surface, while contributions
of macroscopic loops, representing extended states, give
rise to the magnetic moment $\Delta \vec{M}_{a}$
parallel with the cylinder axis.
In the equilibrium $\Delta \vec{M}_{a}$ vanishes because
of the zero total current. Any electron transfer to 
state with opposite velocity direction gives rise to non-zero
current as well as non-zero $\Delta \vec{M}_{a}$. Since
the mass-center positions of different states generally do
not coincide the transfers are leading to non-equilibrium
charge distribution defining an electric field which is in
average perpendicular to the current flow.

The above discussion leads to the conclusion that the Hall
current for the strip of the width $w$ is controlled by only
the part of the magnetic moment defined by Eq.~(\ref{Ma}).
Replacing $\vec{M}$  in Eq.~(\ref{Jdef}) by $\vec{M}_a$ gives the
following expression for the Hall conductance
\begin{eqnarray} \label{G_H}
G_H(\mu) \, \equiv \, \frac{1}{R_H(\mu)} \, = \, - \frac{ec}{w}
\left( \frac{\partial M_a}{\partial \mu} \right)_{\vec{B}}
\, =  \nonumber \\
= \, \frac{e^2}{2 \pi w} \, \sum_{\beta} \int \,
\delta \left( \mu-\epsilon_{\beta}(k) \right)
X_{\beta}(k)  v_{\beta}(k) \, dk
\, =  \nonumber \\
= \, \frac{e^2}{h} \, 
\sum_{j=1}^{N_p} \,
\frac{X_{\beta}(k_{j,\beta}^{+}) - X_{\beta}(k_{j,\beta}^{-})}{w}
\; ,
\end{eqnarray}
where $j$ counts pair states $k_{j,\beta}^{+}$ and
$k_{j,\beta}^{-}$ at the Fermi energy $\mu$ having opposite     
velocity directions and $N_p$ is their number.
The considered boundary conditions
imply that $G_H$ is just equal to the inverse value of the
Hall resistance $R_H$.

The obtained result can be rederived by using
Landauer-B\"{u}ttiker-like view to the electron transport.
The studied finite sample is supposed to be attached via
ideal leads to the
source and drain. Difference of the source and drain
chemical potentials, $\Delta \mu$, induces the
current which depends on the occupation of states within
ideal leads. Assuming  scattering leading
to uniform occupation of the outgoing channels within ideal leads,
represented by transmission and reflection probabilities, $t$ and
$r=1-t$, respectively, we get
\begin{equation} \label{J_Buttik}
J \equiv  \frac{e}{h}  \sum_{j=1}^{N_p} \,
\left[ \Delta \varepsilon_{\beta}(k_{j,\beta}^{+}) - 
r \, \Delta \varepsilon_{\beta}(k_{j,\beta}^{-}) \right]   
= \frac{e}{h} N_p t \Delta \mu \, .
\end{equation}
The energy intervals
$\Delta \varepsilon_{\beta}(k_j^{+})=\Delta \mu$
of incoming states just equals to that of outgoing
states $\Delta \varepsilon_{\beta}(k_j^{+})$.
Denoting chemical potential of the drain as $\mu_0$,
the effective potential within the ideal lead on the
drain side equals to $\mu_0+t \Delta \mu /2$,
while that on source side is $\mu_0+ (1+r) \Delta \mu /2$.
Identifying their difference with the voltage drop
the above outlined standard textbook procedure gives
the following expression for the sample resistance
\cite{textbook}
\begin{equation}
R \; = \; \frac{r \Delta \mu}{e J} \; = \;
\frac{h}{e^2} \, \frac{r}{N_p t}
\; ,
\end{equation}
where the ratio $t/r$ defines the relaxation time.

The current $J$ can alternatively be expressed as a function
of the difference between effective chemical potentials of
the incoming and outgoing channels,
$\mu^{+} \, \equiv \, \mu_0 + \Delta \mu$ and 
$\mu^{-} \, \equiv \, \mu_0 + r \Delta\mu$, respectively.
The expression Eq.~(\ref{J_Buttik}) then obeys
the following form
\begin{eqnarray} \label{J_Buttik_Hall}
J \equiv  \frac{e}{h} \,
\frac{\mu^{+} - \mu^{-}}
{ X^{+} - X^{-} }
\sum_{j=1}^{N_p} \, \left[ 
X_{\beta}(k_{j,\beta}^{+}) - X_{\beta}(k_{j,\beta}^{-})
\right] \, , \\
X^{\pm} \, \equiv \, \frac{1}{N_p} \sum_{j=1}^{N_p}
X_{\beta}(k_{j,\beta}^{\pm})
\, , \nonumber
\end{eqnarray}
where $X^{+}$ and $X^{-}$
denote mean positions of incoming and outgoing states,
respectively. The difference $\mu^{+} - \mu^{-}$
determines the strength of the voltage across the strip
and it is the same at both ideal leads, that close to
the source and that at the drain side.
It implies that the expression
Eq.~(\ref{J_Buttik_Hall}) defines a Hall resistance.
However, for microscopic systems the measured voltage difference
depends not only on the properties of the studied system but also
on the voltage detection techniques. To find the relation
between $\mu^{+} - \mu^{-}$ and measured voltage drop is thus
non-trivial problem.

To proceed further, let us consider a macroscopic system
composed of the parallel microscopic strips. Assuming no
particle transfer between them and a gradient of the
chemical potential perpendicular to the current flow,
conditions for the standard measurement on the Hall
bar samples are ensured.  Constant gradient
gives the same difference $\mu^{+} - \mu^{-}$ within
each of the strips and the substitution
\begin{equation} 
\frac{\mu^{+}-\mu^{-}}
{X^{+} - X^{-}}
\, \rightarrow \, \frac{d \mu}{dx}
\, \rightarrow \, e {\cal{E}}_x \; ,
\end{equation}
immediately gives Eq.~(\ref{G_H}) defining the
Hall resistance of the macroscopic strip.

To illuminate the discussion let us apply the above
presented general results to one particular example, the
periodically modulated system in the strong magnetic field
giving three magnetic flux quanta per unit cell, i.e.   
$p/q=3$. Potential modulation
\begin{equation} \label{Vper}
V_{\rm b}(x,y) \, = \, V_0 \,
\left( \cos Kx \, + \, \cos Ky \right) \; ; \;
K \, \equiv \, \frac{2 \pi}{a} \; ,
\end{equation}
is assumed to be weak, i.e. $V_0$ is much less than
the Landau level spacing $\hbar \omega_c$ ($\omega_c \equiv
eB/m^{\ast}c$), $a$ denotes the lattice constant.
The typical energy dispersion for the lowest Landau level,
which is broadened by weak modulation, is shown in
Fig.~2.

The confining potential $V_{\rm conf}(x)$ was modeled by
$V_c[(x_L-x)/a]^4$ for $x<x_L$, by
$V_c[(x_R-x)/a]^4$ for $x>x_R$, and it was zero for
$x_L<x<x_R$. The region of the periodic potential within
the interval $(x_{iL} , x_{iR})$ has been surrounded by
strips of zero potential that destroy interference
between magnetic edge states located at opposite edges
through the {\it "bulk"} states, the situation expected
in macroscopic systems. The same model has been considered in
the already published paper \cite{Streda94},
where energy spectra as a function of the mass center
are presented.

System boundaries give rise to edge state branches composed
of pair states. Among them magnetic branches are the most
important since states of each pair having opposite velocities
are located at opposite strip edges \cite{Streda94,MacDonald}
and they thus substantially contribute to the magnetic moment.
Magnetic edge-state branches cross energy gaps of infinite systems
and the number of their crossing with the line of the fixed energy
per each edge just equals to the absolute value of the integer $i$
satisfying the Diophantine equation, Eq.~(\ref{Diophantine}).
Whenever $\mu$ cross only magnetic edge branches
the evaluation of the expression Eq.~(\ref{G_H}) gives quantum
Hall values in the limit of the infinite strip width
\cite{Streda94,MacDonald}.

\begin{figure}[h]
\includegraphics[angle=0,width=3.3 in]{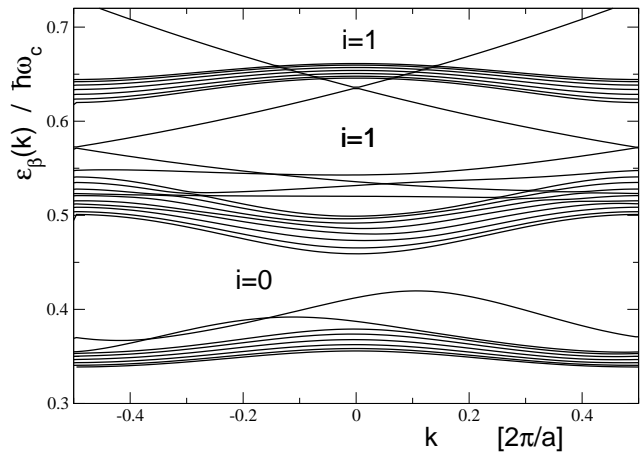}
\caption{Energy spectrum of the lowest broadened Landau level,
$p/q=3$, $V_0=0.2 \cdot \hbar \omega_c$.
Two branches just above the lowest magnetic subband are
composed of non-magnetic edge states while edge state branches
above the central subband are formed by magnetic edge states.}
\label{fig2}
\end{figure}

Energy branches composed of {\it "bulk"} states which are spread
within the strip interior are forming $p=3$ magnetic subbands.
Number of branches increases with rising width of the strip.
To eliminate the details depending on the position of interior
edges, the averaging procedure over the values of $x_{iL}$ and
$x_{iR}$ has been applied to measurable quantities.
By subtracting values obtained for two cases, for which
the width of the periodic parts differs by the lattice constant,
the contributions of the studied quantities per elementary
cell has been obtained.
Numerical calculations show that the averaged quantities
per unit cell are practically independent on the form of
the confining potential and that the increase of the total
strip-width above ten lattice constant does not change
the results, at least for the studied example.
Let us further note, that the briefly
outlined numerical procedure allows evaluation of the
derivatives with respect of the magnetic field since, at
least in principle, it can be applied for any value of $B$.

The averaged value of the total magnetic moment $\vec{M}$
has been established by using the thermodynamic relation,
Eq.~(\ref{Mdef}), with the grand canonical potential having
the following explicit form
\begin{equation} \label{strip-omega}
\langle \Omega \rangle \, = \, \left \langle \sum_{\beta,k}
\Theta \left( \mu - \varepsilon_{\beta}(k) \right) \,
\left [ \varepsilon_{\beta}(k) - \mu \right ] \right \rangle
\; ,
\end{equation}
where angular brackets denote the above described averaging.
The obtained results for the derivative of the magnetic
moment with respect of the chemical potential are shown in
Fig.~3. As expected {\it "bulk"} states give diamagnetic
contributions while the edge state contributions have
paramagnetic character corresponding expected values of
the quantum Hall effect.

\begin{figure}[h]
\includegraphics[angle=0,width=3.3 in]{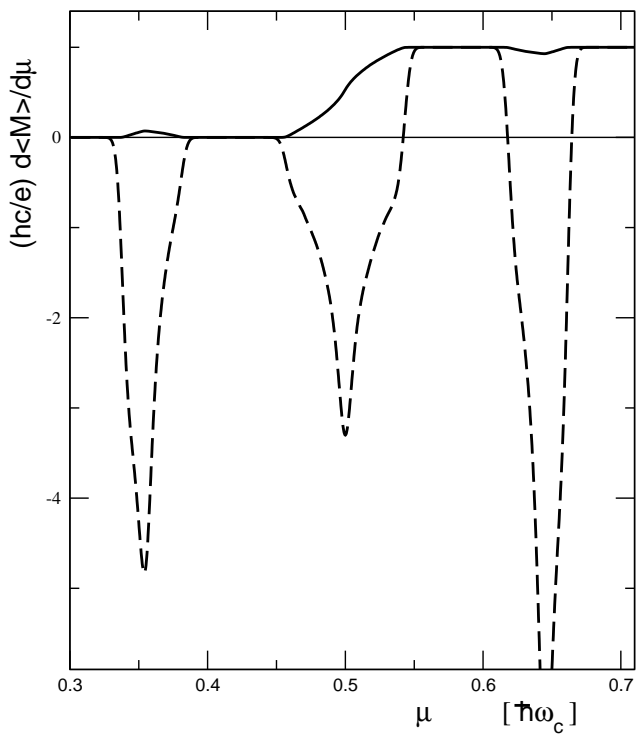}
\caption{Averaged derivatives of the total magnetic moment $M$
(dashed line) and $M_a$ (full line) with respect of the
chemical potential $\mu$ for $p/q=3$,
$V_0=0.2 \dot \hbar \omega_c$ and the strip width around
$16 \cdot a$. The full line coincides with the $\mu$-dependence   
of $-(h/e^2)<G_H>$.}
\label{fig3}
\end{figure}

The averaged Hall conductance $<G_H>$, plotted in Fig.~3 as
a function of $\mu$, has been obtained by
applying the same averaging procedure to the expression
Eq.~(\ref{G_H}). In the regions where only edge states exist
the quantum values of the Hall conductance are reached.
Within the central magnetic subband 
$\langle G_H \rangle$ is approximately determined by the
filling factor of this subband $\nu_c=3\nu-1$ 
[$\nu \epsilon (1/3,2/3)$],
i.e. $\langle G_H \rangle \approx - e^2 \nu_c / h$.
Opposite to the lover subband
giving rise electron-like Hall effect the contribution
of the upper subband has hole-like character.
For higher magnetic fields giving  $p/q=2 n + 1$,
the dependence of the Hall conductance on the energy $\mu$
has qualitatively similar structure. There appear  $n$
electron-like Hall peaks bellow central region and
$n$ dips in the Hall conductance above.
The more rich structures,
to which a separate publication will be devoted,
can be expected for fractions $p/q$ for which $q$ differs
from the unity.

The derived formula, Eq.~(\ref{G_H}), expresses Hall
resistance of the macroscopic systems in terms of Fermi
electron properties. In the presented form its validity
is limited to the systems where the scattering events
lead to the uniform occupation of conducting channels,
by another words if the scattering in macroscopic systems
can be characterized by a single energy-dependent relaxation
time. In principle the same result should be obtained
by using quantum theory of the linear response, i.e.
via the Kubo formula. As has been already mentioned
the results coincide in the case of the non-dissipative
quantum Hall regime \cite{Streda82,Schenker}. It is
also trivial to prove that the same results are obtained
for free electron gas in the weak field limit for which
Landau level quantization is smeared out.
To prove it in the general case is the challenge for
future theoretical studies.

\acknowledgments
{This work was supported by the Grant Agency of the Czech
Republic under Grant No. GACR 202/05/0365.}

\end{document}